# HYBRID PETRI NET MODEL BASED DECISION SUPPORT SYSTEM

**Janetta Culita, Simona Caramihai, Calin Munteanu**

*"Politehnica" University of Bucharest*
*Dept. of Automatic Control and Computer Science*
E-mail: jculita@yahoo.com, sic@ics.pub.ro, mc_aurel@yahoo.com

Abstract: It is well known that the complex system operation requires the use of new scientific tools and computer simulation. This paper presents a modular approach for modeling and analysis of the complex systems (in communication or transport systems area) using Hybrid Petri nets. The performance evaluation of the hybrid model is made by a simulation methodology that allows building up various functioning scenario. A Decision Support System based on the above mentioned methodologies of modeling and analysis will be designed for performance evaluation and time optimization of large scale communication and/or transport systems.

## 1. INTRODUCTION

Complex systems represent a fast growing area of interest for control and optimization. Domains as communication and transport are important fields of interest from this point of view, due to the fact that the present tendency is to build aggregated networks either informatics or of transport (as the European unified railway system project).

Performance evaluation, analysis and optimization of these networks necessitate particular methodologies and formal instruments. Since pure mathematical models are not easy to manage for these categories of systems, decision is often taken on the basis of the simulation of various functioning scenarios. Decision Support Systems (DSS) ensuring the appropriate framework for modeling, analysis and comparison of those scenarios are valuable tools for network managers.

The paper aims to propose a DSS architecture for complex network analysis, based on hybrid Petri Nets modeling and analysis.
Petri Nets (PN) represent a formal tool that was defined in 1962 by Karl Adam Petri, for modeling and analysis of information systems. Their modeling power, capacity of representing concepts as parallelism, synchronization, resource sharing a.s.o. in a clear, intuitive graphical format, have contributed to their wide development and utilization, for very different domains..

More recently, *continuous* PNs were defined (David and Alla, 1987). Autonomous continuous PNs have been shown to be a limit case of discrete PNs. In a continuous PN, the markings, arc weights and firing quantities are non-negative, as in a discrete PN, but are *not necessarily integers*. In a timed continuous PN, maximal speeds are associated with transitions. Other authors have proposed various results concerning these PNs (Demongodin, *et al.*, 2000; Recalde and Silva, 2000; Special issue, 2001).

The initial motivation leading to the concept of continuous PN was an approximate modeling of discrete systems with a large number of states, as a consequence of the management of a large number of entities by the modeled process. By this approach it was possibly to analyze the flow of entities instead of following the evolution of every one. Another domain of application of this type of nets is for continuous system modeling.

Hybrid PNs contain a discrete part and a continuous part, usually interacting (Le Bail, *et al.*, 1991; David and Alla, 2005). Given a timed hybrid PN (timings associated with discrete transitions and maximal speeds associated with continuous transitions), instantaneous behavior is analyzed in the following way: a stable marking of the discrete part is sought, then the instantaneous firing speeds of the continuous transitions are calculated.

The semantics related to instantaneous firing speeds (i.e. local calculation given the marking and feeding flows of the input continuous places of a continuous transition) is relatively easy to define. However, automatic calculation for the whole continuous PN is difficult. Iterative algorithms presented in (David and Alla, 1987) and (Alla and David, 1998) do not work in all cases. Calculation by resolution of a linear programming problem (LPP) was used for some specific cases in (Balduzzi, *et al.*, 2000).

The bases of a speed calculation method for continuous PN are presented in (Munteanu, *et al*, 2005). Many more details are given in (David and Alla, 2004) and software for analysis of Hybrid and Continuous Petri nets is available at http://sirphyco.lag.ensieg.inpg.fr/.

The paper will use the Hybrid Petri Nets (HPN) formalism as the basic approach for designing a Decision Support System for communication and/or transport networks.

The next section will present the specifications of the network analysis problem and the consequent structure and operation of the DSS.

Section 3 will present a short case study, illustrating the methodology of analysis implemented by the DSS. Finally, the conclusion section will present some future research directions.

## 2. DSS STRUCTURE

The block architecture of the Decision Support System is presented in Figure 1. Its structure was designed taking into account the specifications of the problems it has to resolve.

Communication, as well as transport networks, have in common the fact that they have to transfer a certain flow of entities (informational or physical) from one starting point to a destination, via a net of possible paths, which will be called connections, usually with some time constraints. The availability of connections can vary in time, as well as the amount of entities to be transferred and the time limits, so a solution will satisfy only a given set of constraints and will be unique. On the other hand, usually there is searched only a valid solution and not the best one, even if some optimization is desirable.

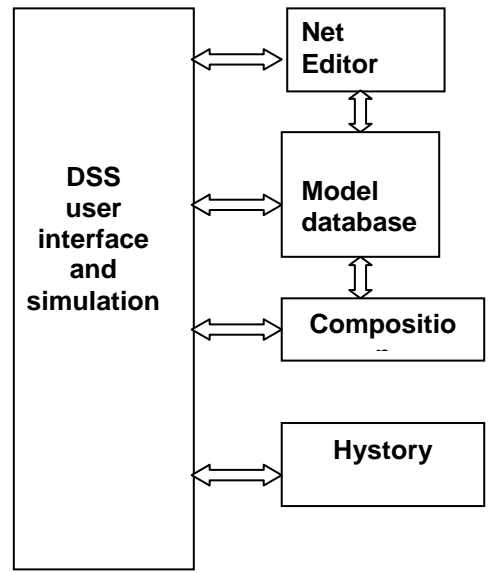

Fig. 1 DSS structure

Since usually the global network is constituted from different sub-networks, there was taken into account the possibility of composing sub-models into a global one, to be analyzed.

Consequently, the DSS will include a net-editor that will allow the user to design models of networks represented in HPN formalism.

The nets can be either directly analyzed or stored in a model database. With the models from the database, the composition block permits the building of larger global nets. Finally, the history module let the user to either store analyzed scenarios or to verify and compare previous analyses.

The user interface module allows the user to initialize the structural models with actual parameters and especially to simulate the net functioning.

The following section will present the model design and analysis methodology on a small communication network – case study.

## 3. THE COMMUNICATION SYSTEM- A CASE STUDY

The considered communication system transfers information between two nodes (node 1 and node 4), source and destination, using any available connection, via the other network nodes (i.e., node 2 and node 3).

The information to be transferred, consisting of a given number of packets, has to reach the destination into a given amount of time. The packets can be sent by different routes, as the destination node has the possibility to order them as necessary for reconstructing the information. Therefore, an efficient distribution of packets on different paths, starting from the source node could improve the overall transmission time. The problem to be analyzed is how to distribute packets on different routes, according to their availability and transfer speed, in order to meet the overall time constraints. Obviously, the routes will not consider twice the same node.

*Hybrid Petri net model*

Figure 2 presents a communication sub-net that aims to transmit a number of information packets from the node 1 to node 4, eventually using internal nodes 2 and 3. Thus, the possible transfer routes are: 1 -> 4, 1 -> 2 -> 4; 1 -> 3 -> 4; 1 -> 2 -> 3 -> 4 and 1 -> 3 -> 2 -> 4 as specified in the figure 2.

As a part of a larger communication system, it is assumed that nodes 1, 2 and 3 perform also other jobs more or less important than the transfer activity. Moreover, for a particular system configuration some physical connections between nodes could be not available. The priorities of jobs and the availability of connections will be modeled by the continuous transitions speeds and priorities.

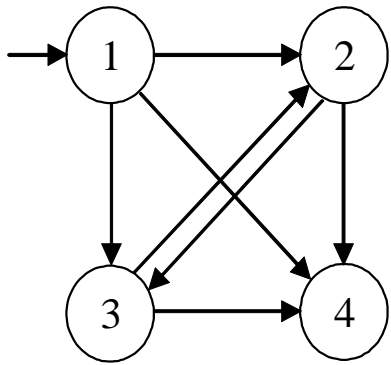

Fig. 2. Communication subsystem

It is assumed that the reader is familiar at least with regular and timed Petri Nets, so as no in-deep explanation for the model will be provided.

The modeling framework is hybrid Petri Nets, as it is suitable for the description of both continuous states or actions (transfer) and discrete states (the existence of a connection). Figure 3 illustrates the hybrid Petri net structural model of the communication sub- system. For simulation purposes the marking will be initialized in order to reflect the associated connection state. It should be noted that even if a connection is available at a certain moment, it could physically breaks-down. The presence/absence of a connection is modeled by the discrete transition - place group.

In figure 3, transitions $T_1 \div T_3$ model the working speed of the nodes $1 \div 3$. Continuous place $P_1$ is associated to node 1. The continuous transitions $T_4 \div T_6$ model the transmission speed from node 1 to nodes 2, 4 and 3 respectively. $T_{15}$ is used for modeling the other jobs the node 1 can execute, other than the transmitting activity, with higher priority. Similarly $T_{16}$ models the supplementary jobs of node 1, other than the transmitting activity, having lower priority. During the simulation of different activity scenarios, speeds and priorities can be modified in order to compare functioning regimes.

Place $P_5$ models the number of packets to be transmitted to the destination. The modeling approaches for both nodes 2 and 3 are similar: $P_2$ models node 2 and $P_3$ models node 3. From these nodes the information could be transmitted towards node 3 or 4 (from node 2), respectively nodes 2 or 4 (from node 3) by transitions $T_7$, $T_8$ and $T_9$ respectively $T_{10}$, $T_{11}$ and $T_{12}$. The transition $T_7$ models the information transmission from node 2 to node 3 and $T_{12}$ from node 3 to node 2. For the route 2 -> 4 transitions $T_8$ and $T_9$ are used that distinguish between the ways 1-> 2- > 4 and 3 -> 2 -> 4. Identically, in order to model the transmission on the route 3 -> 4 $T_{10}$ and $T_{11}$ are used (corresponding to connections 1-> 3 and 2 -> 3).

Places $P_6 \div P_9$ are intermediate and model the information flow transferring from node 1 to node 2 ($P_6$), 1 to 3 ($P_7$), 3 to 2 ($P_8$) and 2 to 3 ($P_9$). The transitions $T_{14}$ and $T_{17}$ model the tasks more important than the transmission of the nodes 2 respectively 3; the transitions $T_{13}$ si $T_{18}$ model the tasks less significant than the transmission of the nodes 2 respectively 3. The place $P_4$ corresponds to node 4. The transfer towards node 4 is already encoded in the transitions $T_5$, $T_8$, $T_9$, $T_{10}$ and $T_{11}$.

All low priority transitions ($T_{13}$, $T_{16}$ and $T_{18}$) will have infinite maximal execution speed so that the places $P_1$, $P_2$ and $P_3$ will not accumulate tokens.

*System analysis (scenario analysis)*

Each transition $T_4 - T_{11}$ is enabled also by a discrete place. By unmarking a discrete place, the absence of the associated connection between two nodes is evidenced. There are two configuration possibilities: one of them is the setting of the discrete marking; the other consists in time association to discrete transitions (temporizations), so as enabling /disabling (of a connection) is realized during the system analysis

Different analyzis scenarios could be obtained by setting the maximal speeds associated to either the transfer activities or other processing jobs. Its values could be constant a priori established, piecewise constant or stochastic (generated by the computer) on time intervals. Also by setting certain priorities/sharing for different possible transfer route various functioning scenario will be constructed.

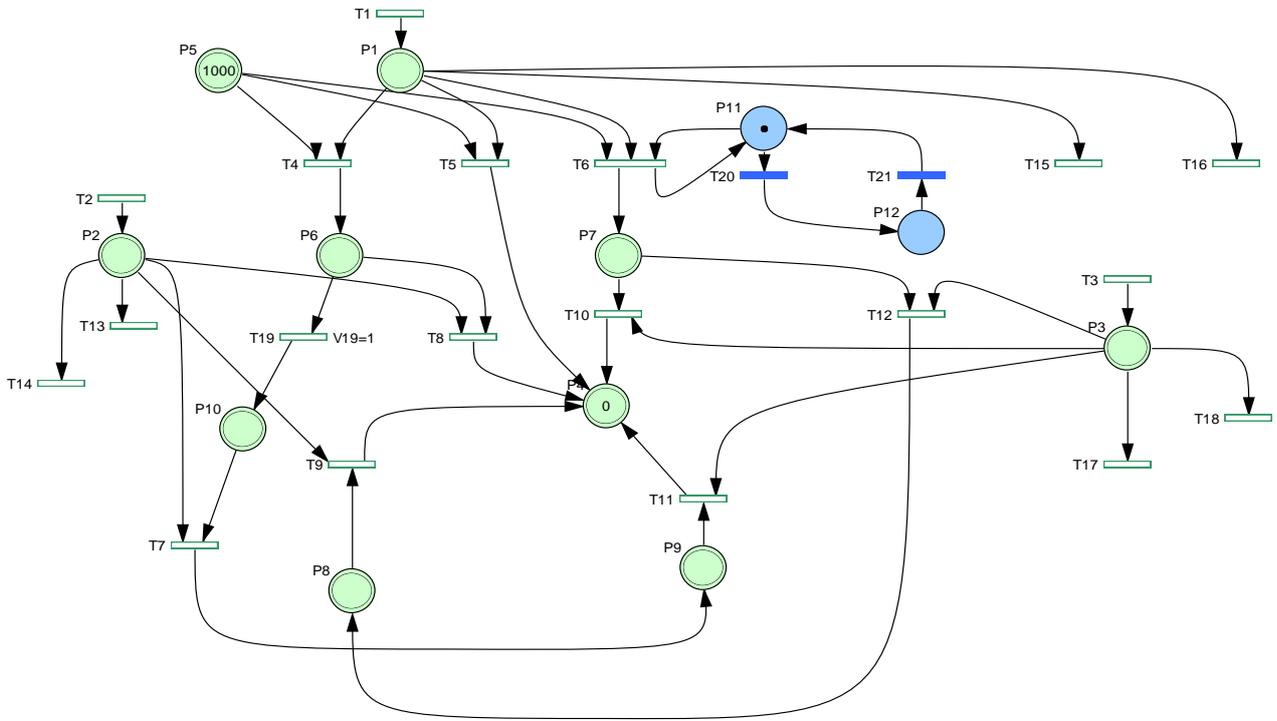

Fig. 3. Hybrid Petri net model

If a certain route is preferred, the associated modeling transition will be associated with higher priority than the other conflicted transitions modeling the other routes. Should no route is favorite (main), all the transitions will have the same weight 1 in a common sharing group. The sharing situation must be carefully chosen, as in the RPC analysis algorithms there are only three priority/sharing combinations (David and Alla, 2005).

The analysis of the communication /transport system assumes the following steps:
- setting the configuration of a communication network (the user sets the initial discrete marking that corresponds to available connections);
- setting the packet transmission speeds on each direct link. This could be done either by the user or by the system. In the last case stochastic values could be associated to the speeds.
- choosing the number of packets composing a message;
- setting the maximum time of message transmission.

Besides these initializations, before the simulation begins, a priority level has to be assigned to each transition.

*Example*
A numerical example will be analyzed by using Syrphico as simulation tool, in order to illustrate the effective speed computing. It is assumed that all connections are available. The continuous marking is set for position P5 (1000). The maximal speed vector is: Vj=[4, 3, 5, 1, 1.5, 2, 1, 0.5, 1, 1, 1, 2, -1, 0.5, 0.5, -1, 1, -1, 1 ] ,

The transition priorities are illustrating in figure 4. Consequently, the instantaneous speed vector results in: V = [4, 3, 5, 1, 1.5, 1, 0.5, 0.5, 1, 0, 0.5, 1, 0.5, 0.5, 0.5, 0, 1, 2.5, 0.5].

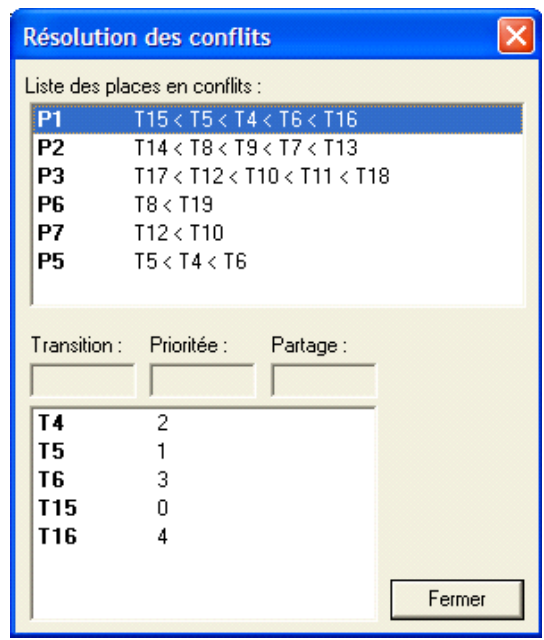

Fig.4. Setting the transition priorities

The marking evolution is then:

$m1(t) = 0+$
$m2(t) = 0$
$m3(t) = 0$
$m4(t) = 3.5*t$

$m5(t) = 1000 - 3.5*t$
$m6(t) = 0+$
$m7(t) = 0+$
$m8(t) = 0+$
$m9(t) = 0+$
$m10(t) = 0+$

From the instantaneous speed vector values, it could be observed that, even if a connection is available, it may not be used due to time consuming server utilization. The other parallel routes realize a transmission time of 286 t.u.

Taking into account two modes of solving the transition conflict, the decision support system will analyze all the possible configurations. The first configuration meeting the time criteria will be selected.

If the time criteria is not accomplished then:
- if there is a place that accumulates markings its input transition will be set with lower priority;
- a larger number of parallel routes are considered;

Before simulation starts the Decision Suport System will chose first (for transitions that model information transmission) the priority level according to the transitions maximal speeds: the transition with biggest maximal speed will have the highest priority level. But this doesn't always lead to best results, as it is shown in the following example:

Suppose that the Petri Net's maximal speeds are: $V = [4, 3, 5, 3, 2, 2, 1, 0.5, 1, 1, 1, 1, \infty, 0.5, 0.5, \infty, 1, \infty]$

Because $T_4$ has the biggest maximal speed (from $T_4$, $T_5$ and $T_6$), it will receive initially the highest priority level (from the three transitions).

*Case A*

Conflict resolutions are: $T_{15} < T_4 < T_5 < T_6 < T_{16}$ (for conflict resolution associated to $P_1$ place) and $T_4 < T_5 < T_6$ (for conflict resolution associated to $P_5$ place). With these maximal speeds and priority levels, the continuous Petri net will have the following evolution:

The first instantaneous transition speed vector will be:

$V = [4, 3, 5, 3, 0.5, 0, 0.5, 0.5, 0, 0, 0.5, 0, 1.5, 0.5, 0.5, 0, 1, 3.5, 0.5]$

Because the input transition of place $P_6$ (i.e. $T_4$ - 3) has an instantaneous speed greater than the sum of maximal speeds for output transitions of $P_6$ (i.e. $T_{19}$ – 0.5 and $T_8$ – 0.5), the markings are accumulated in $P_6$. This means that:
  - node 2 must have a buffer to store the parts that it cannot deliver (due to speed limitations);
  - a second evolution phase is needed in order to deliver the parts from $P_6$. This evolution is characterized by the instantaneous speed vector:

$V = [4, 3, 5, 0, 0, 0, 0.5, 0.5, 0, 0, 0.5, 0, 1.5, 0.5, 0.5, 3.5, 1, 3.5, 0.5]$

First evolution phase will end in 286 time units and the system will send all the packets in 857 time units (which is the end of the second evolution phase).

*Case B*
A lower priority level will be associated to transition $T_4$. Giving to $T_4$ a priority level lower then $T_5$, the conflict resolution rules for places $P_1$ and $P_5$ would be: $T_{15} < T_5 < T_4 < T_6 < T_{16}$ (for $P_1$) and $T_5 < T_4 < T_6$ (for $P_5$).

With these new priority levels, the first phase of the continuous Petri net will be characterized by the following instantaneous transition speed vector:

$V = [4, 3, 5, 1.5, 2, 0, 0.5, 0.5, 0, 0, 0.5, 0, 1.5, 0.5, 0.5, 0, 1, 3.5, 0.5]$

The input transition of $P_6$ will still have an instantaneous speed (i.e. 1.5) bigger than the sum of maximal speeds for output transitions of $P_6$ (i.e. 0.5+0.5=1), so a second evolution phase is needed. This is characterized by:

$V = [4, 3, 5, 0, 0, 0, 0.5, 0.5, 0, 0, 0.5, 0, 1.5, 0.5, 0.5, 3.5, 1, 3.5, 0.5]$

First evolution phase will still end in 286 time units and the system will send all the parts in 429 time units (which is the end of the second evolution phase).

*Case C*

Transition $T_4$ will be set with the lowest priority level. Transition $T_5$ keeps its higher priority and $T_4$ will have a priority level lower then $T_6$; the conflict resolution rules for places $P_1$ and $P_5$ would be: $T_{15} < T_5 < T_6 < T_4 < T_{16}$ (for $P_1$) and $T_5 < T_6 < T_4$ (for $P_5$).

With this new priority levels, the first phase of the continuous Petri net will be characterized by the following instantaneous transition speed vector:
$V = [4, 3, 5, 0, 2, 1.5, 0, 0, 1, 0.5, 0, 1, 1.5, 0.5, 0.5, 0, 1, 2.5, 0]$

This is the only phase needed to transfer all the packets rts from node 1 to node 4. The phase time is 286 time units.

From all three cases it is noticeable that (because of node 1 limitation) only one intermediate node is selected for message transmission and the total number of routes is three.

As it was mentioned before, there is the possibility of increasing the number of routes. This situation appears if neither node 2 nor node 3 could send the packets as fast as they receive them.

CONCLUSION

The paper presents a decision support system constructed on a modular approach for modeling and analysis of the complex systems in communication/transport area. The partial models of intermediate nodes could be composed in order to obtain the whole system model. RPH was chosen as modeling and analysis tool due to a significant modeling power appropriated for complex systems thought as hybrid systems (David si Alla, 2005).

The decision support system inspects the scenarios provided by the system analysis and proposes a time suited solution.

The future research trend is to search the optimal solution corresponding to the minimum transmission time. In this purpose, all the possible situations for priorities/sharing allocation will be analyzed.